\documentclass[epj]{svjour}
\usepackage{floatrow}
\usepackage{sidecap}
\sidecaptionvpos{figure}{t}
\usepackage{graphics}
\usepackage{graphicx}
\usepackage{dcolumn}
\usepackage{bm}
\usepackage{color}
\usepackage{siunitx}
\usepackage{comment}
\usepackage{caption}
\usepackage{subcaption}
\usepackage{adjustbox}
\usepackage{float}
\usepackage{changepage}
\DeclareSIUnit[number-unit-product = {}]\sig{~\sigma}

\newcommand{\st}{\texttt{SixTrack} }
\newcommand{\mad}{\texttt{MAD-X} }
\newcommand{\fl}{\texttt{FLUKA} }
\RequirePackage[colorlinks,citecolor=blue,urlcolor=blue,linkcolor=blue]{hyperref}

\begin{document}
\title{Crystal Collimation Cleaning Measurements with 6.5~TeV protons in the LHC}
\author{Roberto Rossi\inst{1,2} \and Gianluca Cavoto\inst{3} \and Daniele Mirarchi\inst{1} \and Stefano Redaelli\inst{1} \and Walter Scandale\inst{1,2}
 }                 
%
%
\institute{ European Organization for Nuclear Research CERN, Geneva, Switzerland \and Blackett Laboratory, Imperial College, London SW7 2AZ, UK \and Instituto Nazionale Fisica Nucleare INFN, Sezione Roma I, Rome, Italy}
\date{Received: date / Revised version: date}
%
\abstract{
Safe disposal of beam halo is a fundamental requirement of modern superconductive hadron colliders to reduce thermal load on magnets and background to experimental detectors. 
In the CERN Large Hadron Collider (LHC) a multistage system fully compliant with the needs of the baseline operation was build.
At a later stage, two short bent crystals were interleaved to the devices for betatron collimation
to investigate efficiency enhancement of the halo disposal when inserting them as primary stages of the collimation hierarchy.
Each crystal was mounted on a high--accuracy angular actuator, called goniometer, and installed in the clockwise Beam 1, one for the horizontal and one for the vertical plane.
In this paper, measurements of the cleaning performance at collision energy with and without inserting crystals in the standard collimation schemes are discussed; the results are compared to theoretical expectations.
\PACS{
      {Channeling}{}   \and
      {Collimators}{}
     } 
} 

\maketitle

\section{Introduction}\label{intro}

In modern hadron colliders, high beam currents circulating in superconducting magnets are required to operate at the highest energy and luminosity for a large physics potential discovery. 
Halo particles with diffusive behaviour and high probability of being erratically lost along the vacuum pipe should be safely disposed of into room temperature absorbers, to avoid abrupt transition to normal-conducting state of the superconducting magnet coils and to mitigate undesired background in the experimental detectors.
A sophisticated collimation system is generally required to safely clean the circulating beam from the potentially dangerous halo particles.
The LHC collimation system was conceived with a multi--stage configuration based on more than \num{100} collimators, each with two massive amorphous jaws, made of different materials and lengths along the beam direction. 
The jaws are organized in a hierarchical order to progressively dispose the halo particles while reducing their transverse density~\cite{assman}. 
Its performance was demonstrated to fully comply with the requirement of the LHC operation~\cite{valentino2016performance}.

Bent crystals can coherently deflect positive charged particles when planar channeling (CH) is achieved~\cite{tsyganov1976fermilab}. 
Particles with trajectories almost parallel to the crystal planes at the crystal entry face can be captured in channeling regime. 
Particles in channeling are trapped into the electrostatic potential well, generated in highly--ordered lattice material. 
They will follow the plane curvature for moderate values of the bending radius with quasi-harmonic trajectories, perturbed by multiple Coulomb scattering with nuclei and electrons of the atomic planes. During the crystal traversal, the trajectory deflection of a channeled particle is equal to the crystal bending angle. Crystal-particle interactions are discussed in details in~\cite{biryukov,taratin}.

Applications of crystal assisted beam manipulation in modern colliders are recalled in~\cite{scandale2012use}. 
In particular, a crystal collimation system should be based on a short bent silicon crystal replacing the primary stage of the standard collimation system. 
The expected advantages are numerous. 
The crystal deflection of the halo particles is larger than that achievable due to other techniques (e.g. multiple Coulomb scattering in amorphous materials). 
The secondary stage aperture can thus be larger and the induced beam impedance smaller. 
Finally, the debris due to nuclear interactions and the diffractive losses are expected to be considerably reduced during the crystal traversal. 
Attention should however be payed to the design of the secondary stages that should sustain a considerably increased loss density distribution while ensuring a minimum leakage power loads.
In the LHC, crystals bent by mechanical stress can be used to achieve sufficient deflection (up to several tens of \si{\micro\radian}) to steer the particles into a single absorber. 
A layout for crystal collimation in LHC was conceived~\cite{Assmann2006optics,dan_layout} and installed during the first Long Shutdown (LS1) of LHC runs, initially consisting of two crystals, one per transverse plane, in the clockwise beam direction Beam 1 (B1).
With this test--bench channeling has been observed at the record energy of \SI{6.5}{\tera\electronvolt} with both proton~\cite{lhc_ch} and lead ion beams~\cite{rossi_phd}. In this paper, we present performance evaluations of the crystal collimation system in LHC and a comparison of experimental results with computer simulations.

\section{LHC Collimation and Performance Evaluation}\label{sec:lossmaps}

In the LHC design, two insertions are dedicated to collimation, one for transverse and one for off-momentum beam halo cleaning. 
This choice ensures optimal performance at high-luminosity operation. 
The IR7 system underwent an important upgrade in LS2~\cite{redaelli2020collimation}, however here we refer to the Run 2 layout~\cite{martinez2019run}.
Hereafter we will concentrate on the configuration of the so-called betatron collimation insertion, located in the Interaction Region 7 (IR7) which houses three set of collimator families: the primary collimators (TCP), used as the first stage of the system; the secondary collimators (TCSG), used as a secondary stage to dilute the density of the particles escaped from the primary collimators; the active absrobers (TCLA), used to intercept high-amplitude showers. 
The hierarchy is established in order to dispose the beam halo as well as the particle showers that are produced during the interaction with the collimators themselves. 
TCPCs and TCSGs are made of lower Z materials (graphite composite), while TCLAs are made of higher Z materials (tungsten). 
The former are at a closer aperture to the circulating beam; thus to reduce the beam impedance a low density material is chosen.

The evaluation of the cleaning performance is based on the so-called Loss Map (LM) that provides the loss distribution along the entire ring. 
About \num{4000} ionisation chambers distributed along the \SI{27}{\kilo\meter} ring, used as beam loss monitors (BLM)~\cite{blm}, provide the necessary information to built LMs. 
To evaluate loss patterns induced by the collimation system, a white noise excitation is applied to the circulating beam that induces controlled beam losses~\cite{ADT} and clear signals in all the operational BLMs. 
During controlled beam excitation, the highest loss rate is observed at the primary collimators (TCP), where the diffusing particles encounter the first massive obstacles. 
Thus, the signal of the BLM next to a TCP is proportional to the number of halo particles intercepted. 
The ratio $\eta_i=$counts(BLM$_i$)/counts(BLM$_{TCP}$), evaluated at every machine position $i$, provides a direct measurement of the local value of the collimation inefficiency $\eta_i$ .
The Dispersion Suppressor of IR7 (IR7--DS) is the point where the highest losses on cold magnets are observed; this region is located downstream IR7 and houses the first set of superconducting dipole.
The rising dispersion function in this region causes off-momentum particles to be lost on the machine physical aperture.  
This is the reason why the collimation performance in that specific region is of the highest importance when comparing different collimation scenarios.

\begin{figure*}
  \centering
  \includegraphics[width=0.8\linewidth]{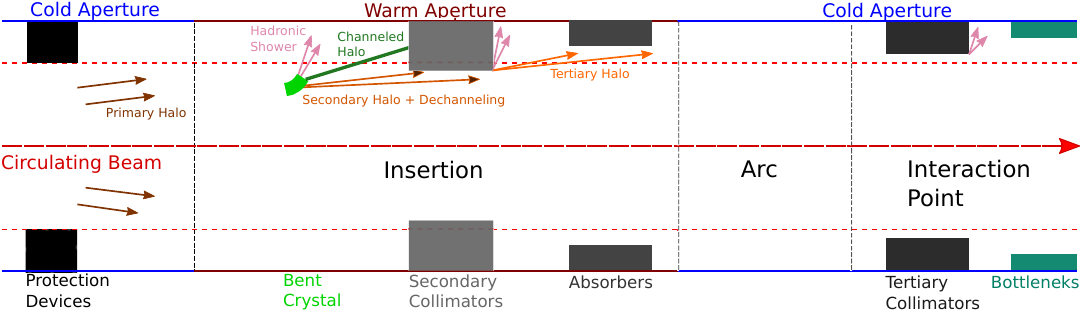}
  \caption{Illustrative view of the crystal collimation system integrated in the betatron collimation insertion of the LHC.}
  \label{fig_scheme}
\end{figure*}

\begin{SCfigure}
  \centering
  \includegraphics[width=0.65\linewidth]{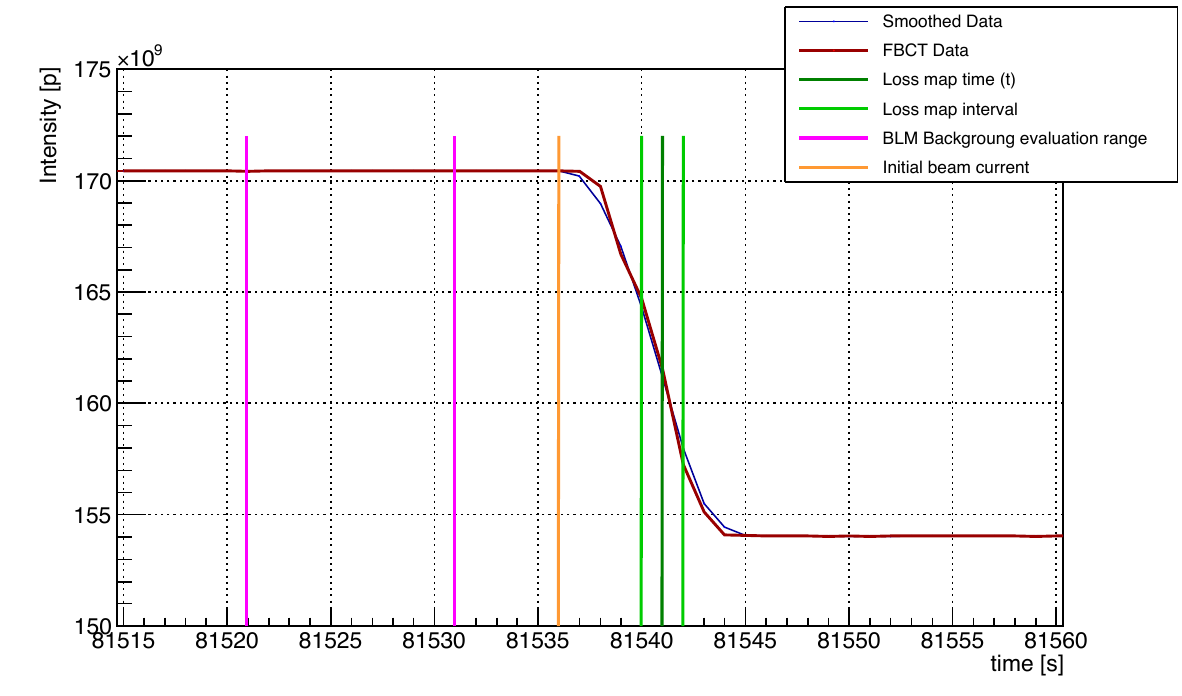}
  \caption{\protect\rule{0ex}{5ex}Beam current as a function of time (solid red). The smoothing process is shown and superimposed (solid blue). The region of steady lifetime is used to evaluate the BLM background (between the magenta lines). The instant when the loss maps is measured (dark green) and the time interval (light green) ($dt$), used for the flux evaluation, are shown. The initial beam current ($I_0$) is shown by the orange line.
    }
  \label{fig:flux}
\end{SCfigure}

The experimental layout for crystal collimation is described in~\cite{dan_layout,rossi_phd,dech_lhc}.
It is based on crystals, mounted in particular devices, called TCPC, able to set the crystal transverse position and angular orientation. 
A secondary collimator is used as the first absorber of the channeled halo, as shown in Fig.~\ref{fig_scheme}.
Given the angular acceptance in the order of \si{\micro\radian} in the energy range of interest, the latter is particularly demanding and required a special interferometer-based goniometer~\cite{gonio1,gonio2}.

\begin{SCfigure}[][t]
  \centering
  \includegraphics[width=0.65\linewidth]{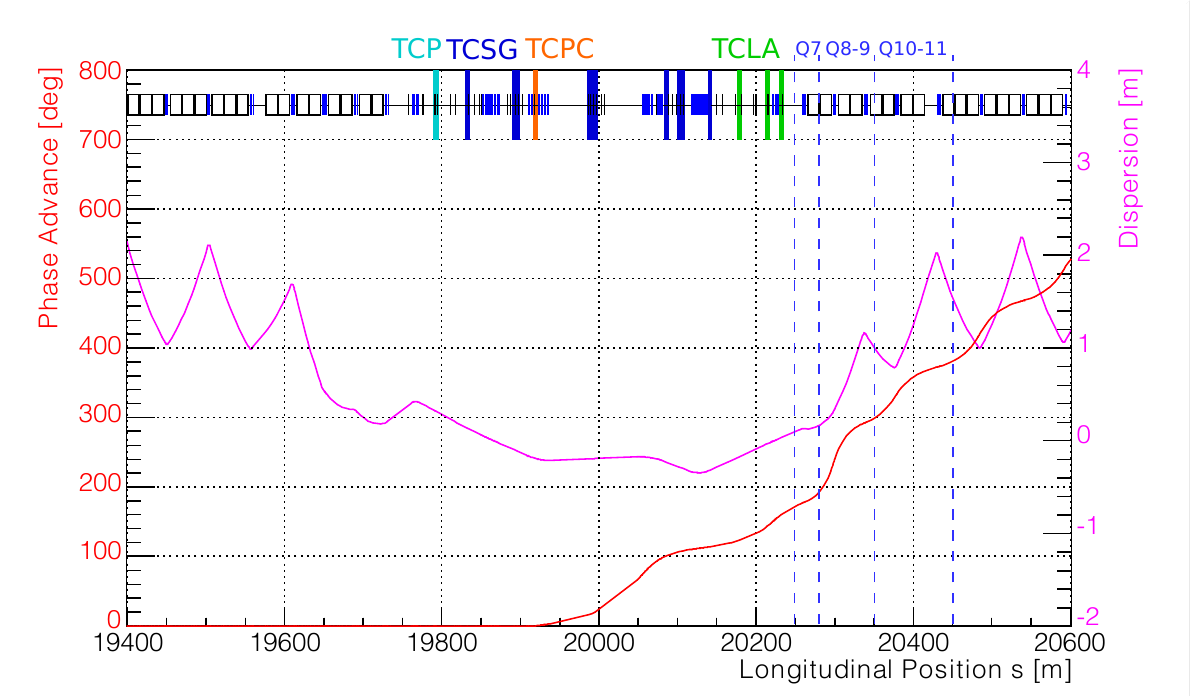}
  \caption{\protect\rule{0ex}{5ex}Dispersion function (solid magenta line) and phase advance (solid red line) with respect to the horizontal B1 crystal around IR7 in LHC. The longitudinal location of standard collimators, the crystals and the regions where the performance are compared are highlighted on top of the plot.
    }
  \label{fig:disp_adv_ir7}
\end{SCfigure}

Particles interacting with a crystal in optimal orientation for channeling produce a flux of debris considerably different from that of an amorphous primary collimator. 
In crystal collimation, the location where the highest losses are expected is not at the crystal itself but at the first collimator used to intercept the channeled halo. 
The crystal is weakly interacting with the incoming beam, whilst the first collimator is irradiated by the large flux of particles deflected coherently by the crystal.
This implies that the loss signal downstream the crystal is not proportional to the number of particles intercepted by the primary collimator, thus, it cannot be used to normalize the beam losses along the ring and to eventually compute local collimation inefficiencies.
To compare the performance of the crystal collimation to the standard one, a different normalization procedure of the LM has been proposed~\cite{lhc_ch}. 
Each BLM signal is normalised to the flux of particles that are lost from the beam during the excitation, evaluated from the bunch--by--bunch beam current measurements~\cite{Belohrad2010Current}, as shown in the example in Fig.~\ref{fig:flux}.
Normalising the loss patterns by the beam loss flux, e.g. producing a pattern of counts(BLM$_i$)/(p/s), allows a direct comparison between the standard and crystal-based systems.


\begin{table*}[t!]
\caption{IR7 collimators positions (in \si{\sig} units) during flat top loss maps measurements with proton beams. Collimators are listed by descending order in longitudinal position.}
\label{tab:coll_p}
\centering
\begin{tabular}{cc  c  cccccc  ccc}
\hline\noalign{\smallskip}
\multicolumn{2} {c}{Collimators}  	    &Standard	    & \multicolumn{6} {c} {Horizontal}			            & \multicolumn{3} {c} {Vertical} \\
{Name} & {Type}  	                    &Reference {[\si{\sig}]}	& \multicolumn{6} {c} {Crystal [\si{\sig}]}			    & \multicolumn{3} {c} {Crystal [\si{\sig}]}\\

{}	& {}	                                & {}	&{1}	&{2}	&{3}	&{4} 	&{5}	&{6}		    &{1}	&{2}	&{3\text{*}}	\\

\hline\noalign{\smallskip}
TCPs			&primary                &\num{5.5}	&\num{7.5}	&Out 	    &\num{7.5}	&\num{7.5}	&\num{7.5}	&Out	    & Out	    & Out	    & Out	\\
TCSG.A6L7	    &secondary skew         &\num{7.5}	&\num{7.5}	&Out 	    &\num{7.5}	&\num{7.5}	&\num{7.5}	&Out		& Out	    & Out	    & Out	\\
TCPCV.A6L7	    &vertical crystal       & Out		& Out		&Out	    &Out	 	& Out		&Out    	&Out 	    &\num{5.5}	&\num{5.5}	&\num{5.5}	\\
TCSG.B5L7	    &secondary skew         &\num{7.5}	&\num{7.5}	&Out	    &\num{7.5}	&\num{7.5}	&\num{7.5}	&Out	    & Out   	& Out	    & Out	\\
TCSG.A5L7	    &secondary skew         &\num{7.5}	&\num{7.5}	&Out		&\num{7.5}	&\num{7.5}	&\num{7.5}	&Out	    & Out	    & Out	    & Out	\\
TCSG.D4L7	    &secondary vertical     &\num{7.5}	&\num{7.5}	&Out		&\num{7.5}	&\num{7.5}	&\num{7.5}	&Out	    &\num{7.5}	&\num{7.5}	&\num{7.5}	\\
TCPCH.A4L7	    &horizontal crystal     & Out		&\num{5.5}	&\num{5.5}	&\num{5.5}	&\num{5.5}	&\num{5.5}	&\num{5.5}	& Out	    & Out	    & Out	\\
TCSG.B4L7	    &secondary horizontal   &\num{7.5}	&\num{7.5}	&\num{7.5}	&\num{7.5}	&\num{7.5}	& Out	    & Out	    &\num{7.5}	& Out	    &\num{7.5}	\\
TCSG.A4L7	    &secondary skew         &\num{7.5}	&\num{7.5}	&\num{7.5}	&\num{7.5}	& Out		& Out	    & Out	    &\num{7.5}	& Out	    &\num{7.5}	\\
TCSG.A4R7	    &secondary skew         &\num{7.5}	&\num{7.5}	&\num{7.5}	&\num{7.5}	& Out		& Out	    & Out	    &\num{7.5}	& Out	    &\num{7.5}	\\
TCSG.B5R7	    &secondary skew         &\num{7.5}	&\num{7.5}	&\num{7.5}	& Out		& Out		& Out	    & Out	    &\num{7.5}	& Out	    &\num{7.5}	\\
TCSG.D5R7	    &secondary skew         &\num{7.5}	&\num{7.5}	&\num{7.5}	& Out		& Out		& Out	    & Out	    &\num{7.5}	& Out	    &\num{7.5}	\\
TCSG.E5R7	    &secondary skew         &\num{7.5}	&\num{7.5}	&\num{7.5}	& Out		& Out		& Out	    & Out	    &\num{7.5}	& Out	    &\num{7.5}	\\
TCSG.6R7		&secondary horizontal   &\num{7.5}	&\num{7.5}	&\num{7.5}	&\num{7.5}	&\num{7.5}	&\num{7.5}	&\num{7.5}	&\num{7.5}	& Out	    &\num{7.5}	\\
TCLAs	        &shower absorbers       &\num{11.0}	&\num{11.0}	&\num{11.0}	&\num{11.0}	&\num{11.0}	&\num{11.0}	&\num{11.0}	& \num{11.0}	& \num{11.0}	& \num{11.0}	\\
\hline\noalign{\smallskip}
\end{tabular}
\footnotesize{\text{*}Crystal oriented in amorphous.}
\end{table*}

In order to compare systematically the critical loss locations, the \emph{leakage factor} is defined: for any LMs it represents the highest normalised loss value observed in a given portion of the accelerator.
For betatron cleaning, the region the IR7--DS is the one where the highest leakage is observed. 
By sub-dividing distinct sectors in this region one can investigate differences in performance between the two collimation systems. 
In particular, three regions in the dispersion suppressor of IR7, are selected.
These are named after the cold quadrupole (Q) that they enclose.
The first one, located at the very end of the straight section, is called Q7.
Even though the dispersion is small the close TCLAs (tungsten absorber) may cause hadron showers in this region if the load of particles impinging on them is very high.
The other two regions are located where the first two peaks in the dispersion function appear.
One is identified by the magnets Q8 and Q9, and the other by the magnets Q10 and Q11.
The dispersion function and the phase advance with respect to the crystal location, are shown in Fig. \ref{fig:disp_adv_ir7} for the horizontal case.
Although not mentioned explicitly is all analyses of the next section, for every loss maps it is explicitly verified that no other critical loss location arise, for any configuration, around the ring.


Two other regions with high loss regime are considered to evaluate the collimation performance: the momentum cleaning insertion (IR3) and near the collimators protecting the LHC dump region (IR6).

\section{Configurations Tested and Experimental Results}\label{methods}

The crystal collimation LMs were measured after aligning the crystals as primary obstacles and orienting them to their optimal channeling direction.
A reduced set of the collimators available in the betatron collimation insertion were used during the crystal collimation measurements.
The various collimator settings employed are described in Tab.~\ref{tab:coll_p}. 
The different arrangements are used to compare the effect of different collimators on the crystal system, as reported in~\cite{rossi_phd}.
For instance, a comparison between configuration \#1 (Cfg\#1) and \#2 in the horizontal plane, highlights the effect of the insertion of collimators upstream the crystal. 
The same information can be gathered with Cfg\#5 and \#6.
Instead, Cfg\#3, \#4 and \#5 show how removing secondary collimators to absorb the residual deflected halo affects the system performance, while Cfg\#6 is the most extreme configuration, where only the last horizontal TCSG is used as absorber.
In the vertical plane, Cfg\#1 and \#2 bring out again the difference between using or not the full set of TCSGs downstream the first one.
And finally, the vertical Cfg\#3 is used to study the effect of using a crystal collimation system with a crystal oriented in amorphous, mimicking a scenario where optimum angular control is lost in operation.

\begin{figure}[htp!]
\centering

\begin{subfigure}{.45\textwidth}
\centering
\includegraphics[width=\linewidth]{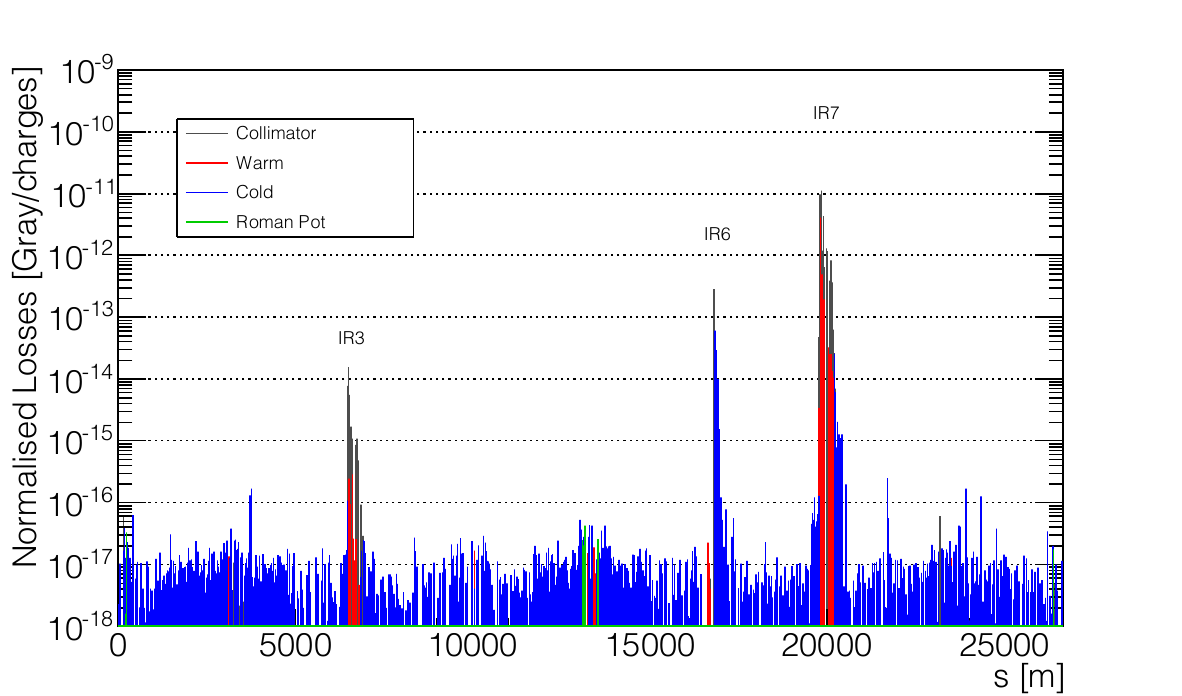}
\caption{Experimental standard system, full ring.}\label{subfig:lm_std_h}
\end{subfigure} %
\begin{subfigure}{.45\textwidth}
\centering
\includegraphics[width=\linewidth]{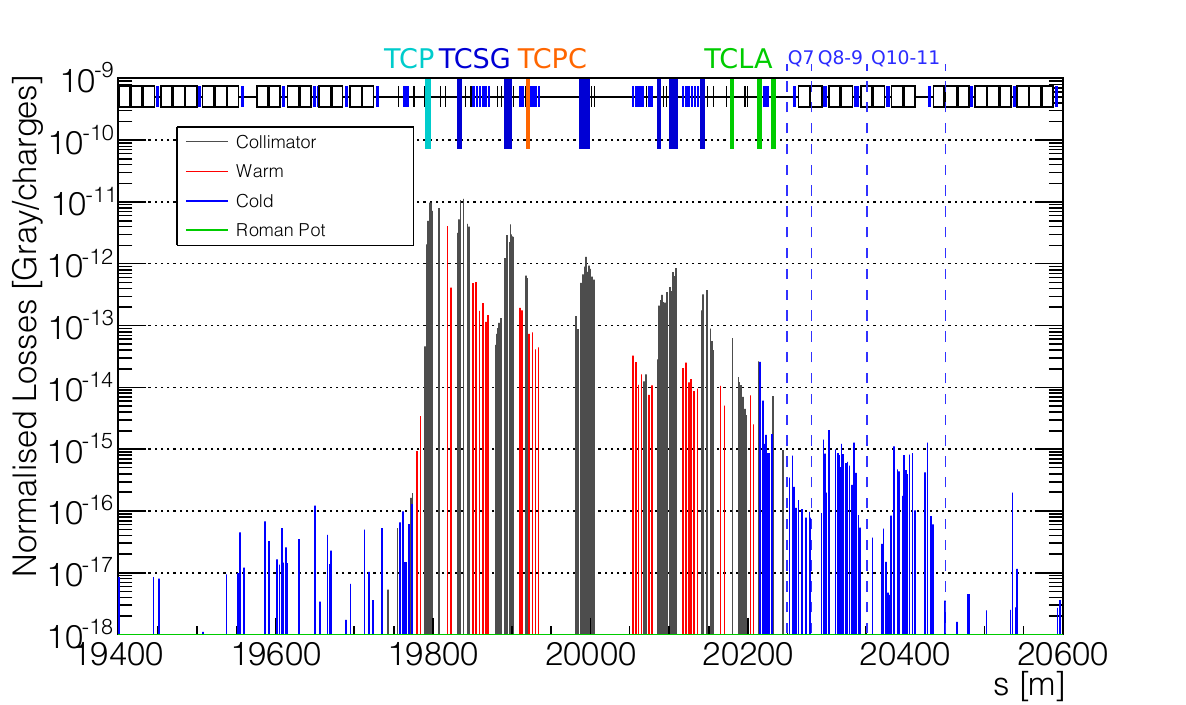}
\caption{Experimental standard system, IR7 zoom.}\label{subfig:lm_std_h_ir7}
\end{subfigure}

\begin{adjustwidth}{0.4cm}{}
\begin{subfigure}{.45\textwidth}
\centering
\includegraphics[width=\linewidth]{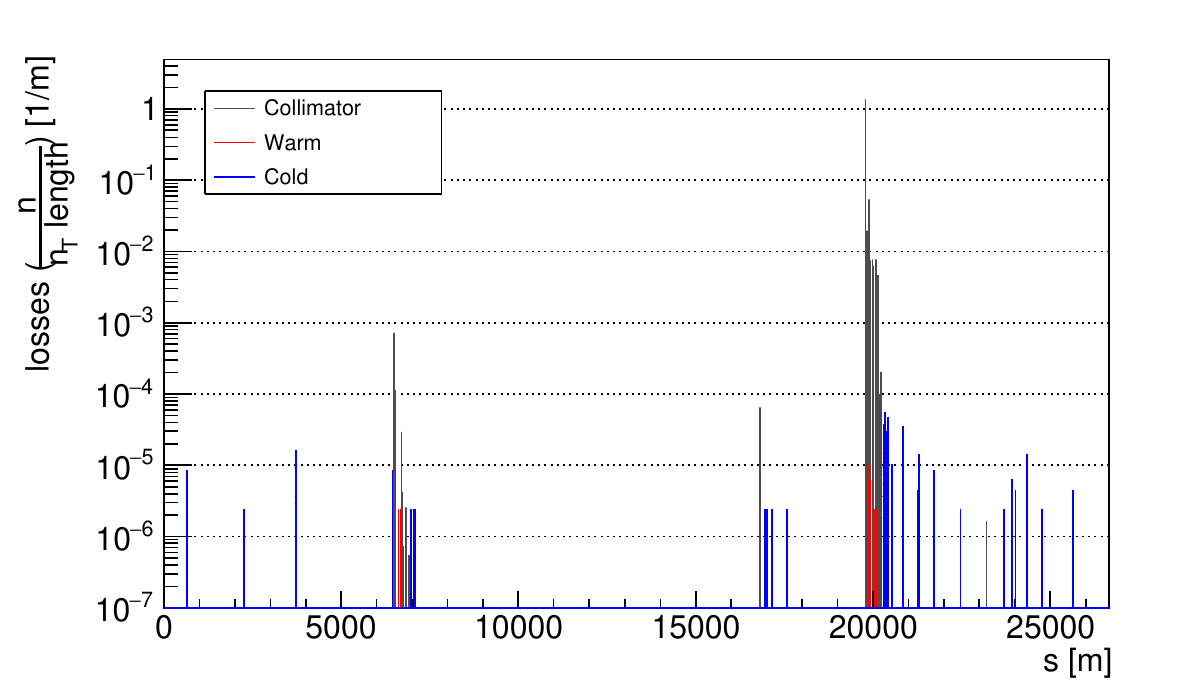}
\caption{Simulated standard system, full ring.}\label{subfig:lm_std_h_sim}
\end{subfigure} %
\begin{subfigure}{.45\textwidth}
\centering
\includegraphics[width=\linewidth]{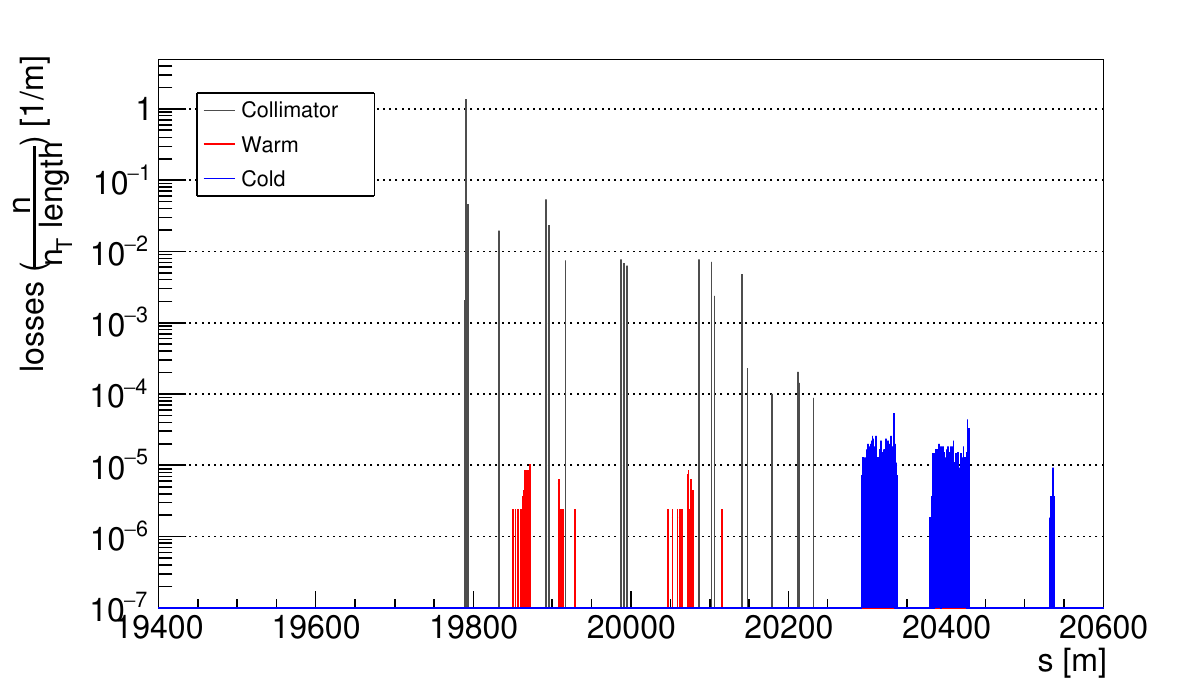}
\caption{Simulated standard system, IR7 zoom.}\label{subfig:lm_std_h_ir7_sim}
\end{subfigure} %
\end{adjustwidth}

\begin{subfigure}{.45\textwidth}
\centering
\includegraphics[width=\linewidth]{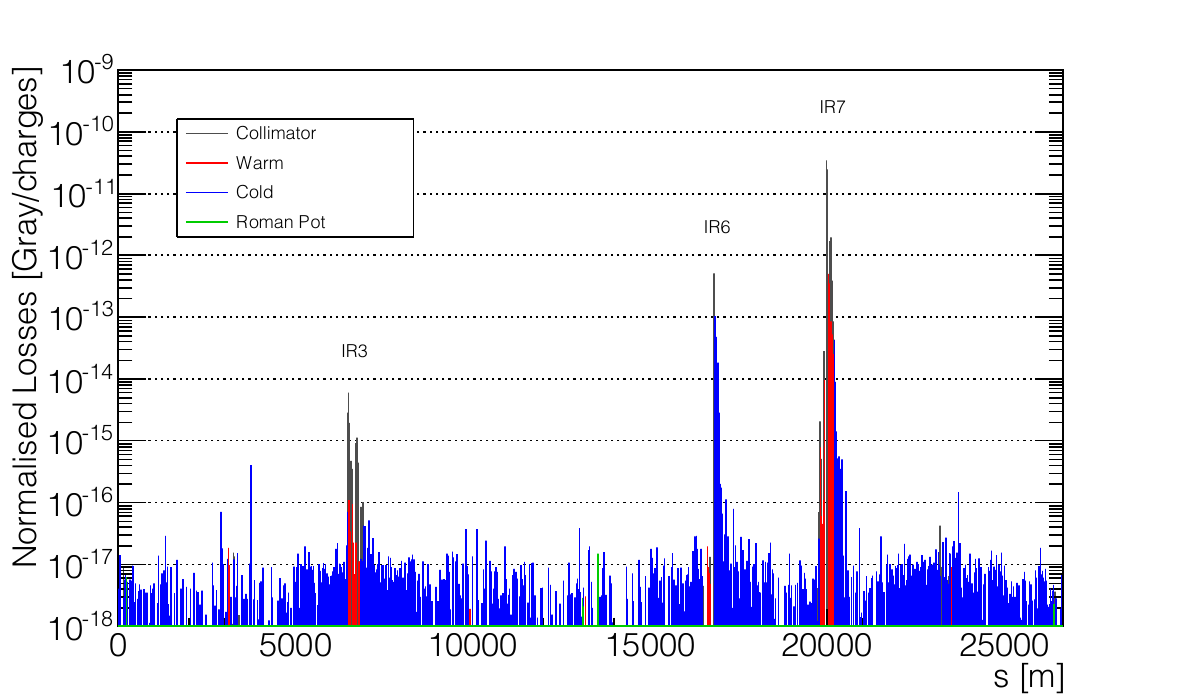}
\caption{Experimental crystal collimation, full ring.}\label{subfig:lm_cry_h}
\end{subfigure} %
\begin{subfigure}{.45\textwidth}
\centering
\includegraphics[width=\linewidth]{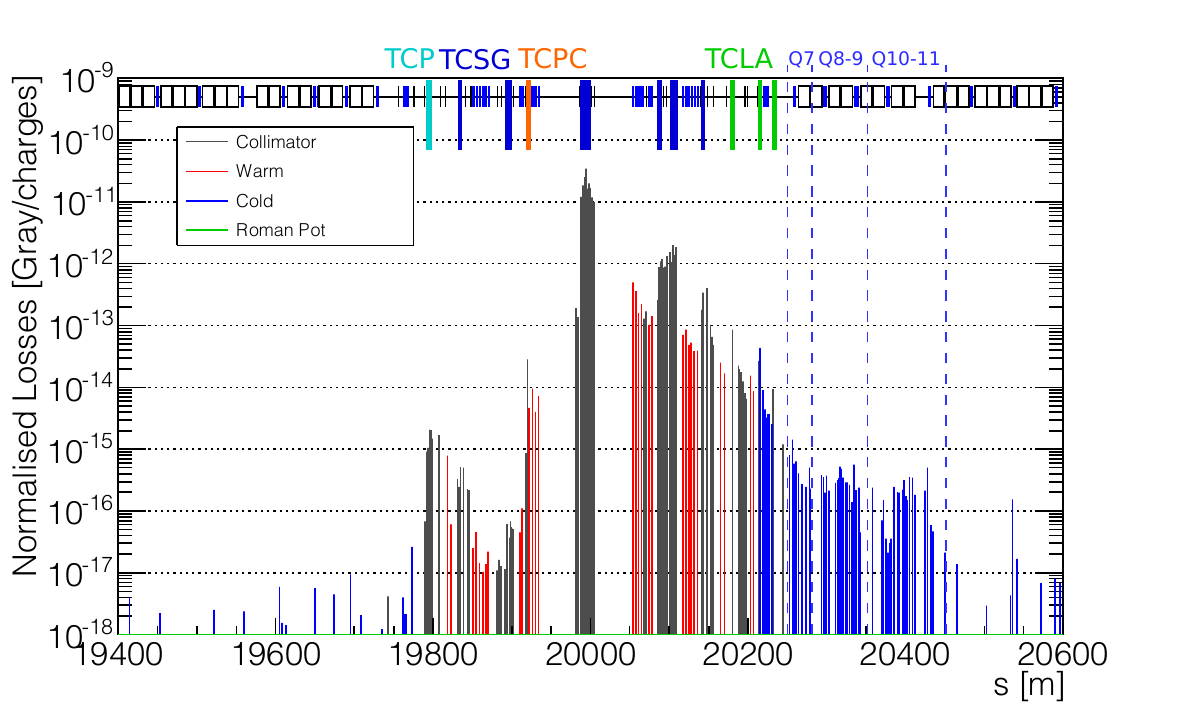}
\caption{Experimental crystal collimation, IR7 zoom.}\label{subfig:lm_cry_h_ir7}
\end{subfigure} %

\begin{adjustwidth}{0.4cm}{}
\begin{subfigure}{.45\textwidth}
\centering
\includegraphics[width=\linewidth]{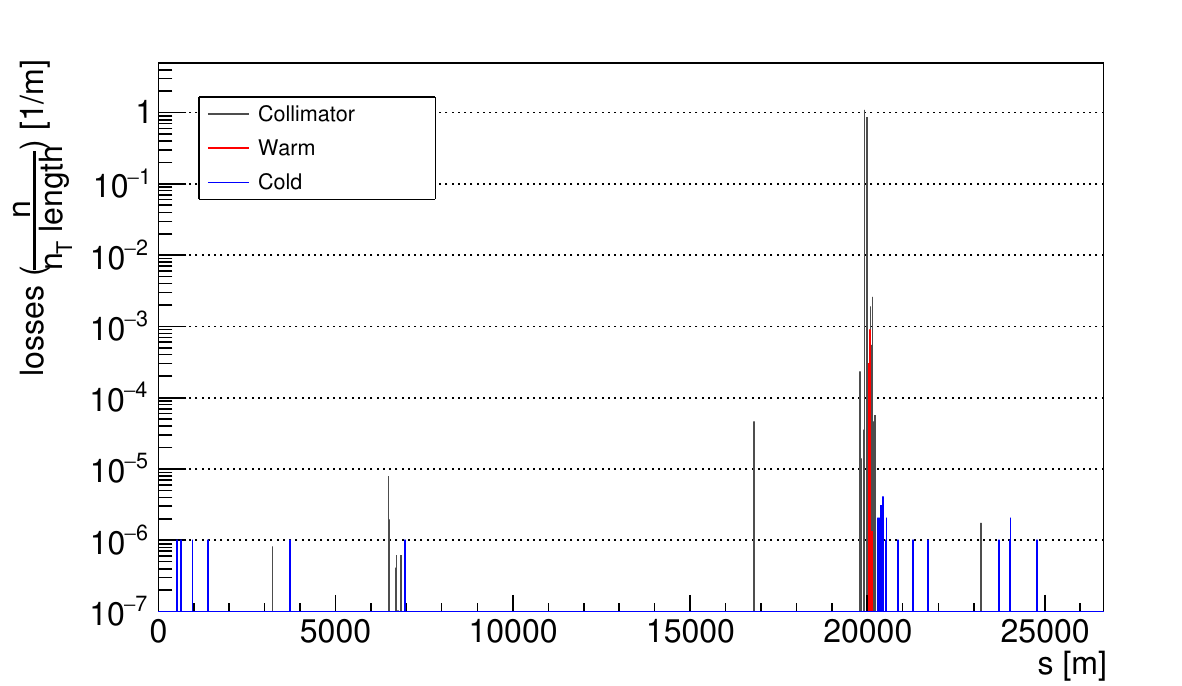}
\caption{Simulated crystal collimation, full ring.}\label{subfig:lm_cry_h_sim}
\end{subfigure} %
\begin{subfigure}{.45\textwidth}
\centering
\includegraphics[width=\linewidth]{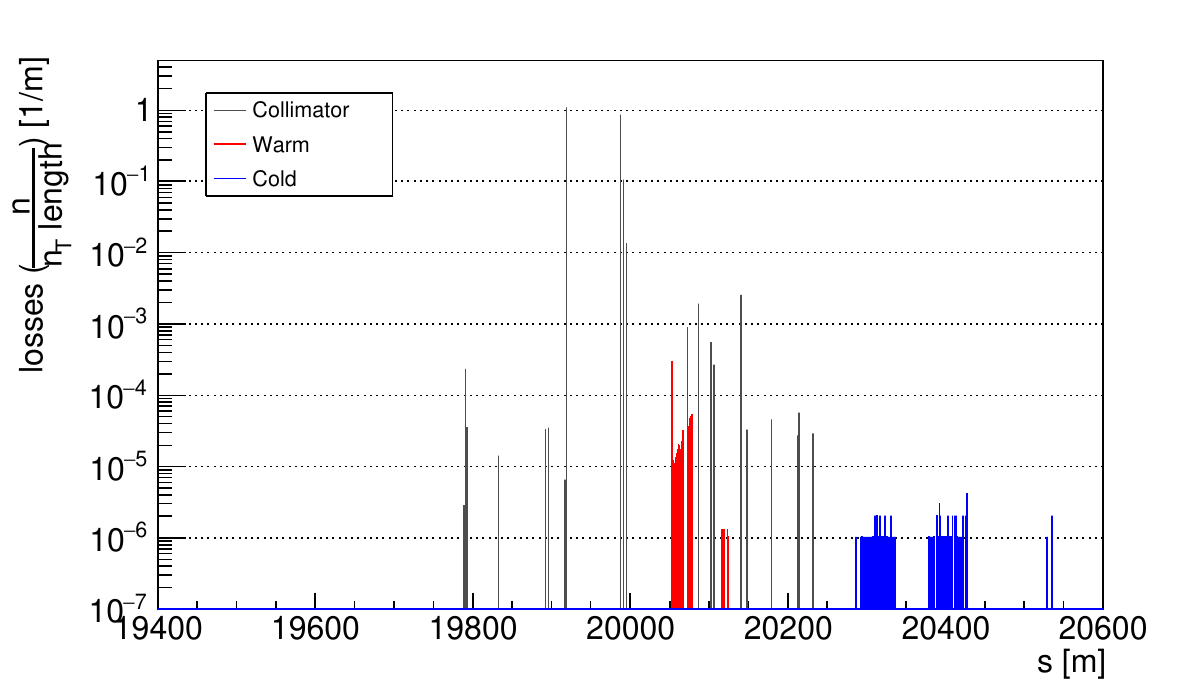}
\caption{Simulated crystal collimation, IR7 zoom.}\label{subfig:lm_cry_h_ir7_sim}
\end{subfigure} 
\end{adjustwidth}
\caption{Horizontal experimental loss maps in the full LHC ring \ref{subfig:lm_std_h} and \ref{subfig:lm_cry_h}, and in IR7 \ref{subfig:lm_std_h_ir7} and \ref{subfig:lm_cry_h_ir7}, with proton beam at top energy. Crystal collimation is used with Cfg\#1 from Tab.~\ref{tab:coll_p}. BLM signals are normalised to the instantaneous beam flux. The simulated loss patterns are shown for for the full ring \ref{subfig:lm_std_h_sim} and \ref{subfig:lm_cry_h_sim}, and in IR7 \ref{subfig:lm_std_h_ir7_sim} and \ref{subfig:lm_cry_h_ir7_sim}, for the respective experimental cases.}
\label{fig:b1_lmch_p}

\end{figure}

The LMs with the standard collimation system were measured to compare the performance of the two system; an example is shown in Figs.~\ref{subfig:lm_std_h} and~\ref{subfig:lm_std_h_ir7}. 
In the standard collimation, the highest loss peak is observed at primary collimators position (see Fig.~\ref{subfig:lm_std_h_ir7}). The LMs obtained with the crystal collimation system in the full machine and in IR7, are shown in Figs.~\ref{fig:b1_lmch_p} (\ref{subfig:lm_cry_h} and \ref{subfig:lm_cry_h_ir7}, respectively) for the horizontal plane in Configuration \#1 (Cfg\#1), defined in Tab.~\ref{tab:coll_p}. 
Due to the reduction of nuclear interaction that occurs when crystal is oriented in channeling, losses are lower than the primary stage of the standard system.
Instead, as shown in Fig.~\ref{subfig:lm_cry_h_ir7}, the highest loss peak is observed at the location of the secondary collimator used as absorber.




\section{Simulations Tools}\label{sim}

Simulation tools are used to reproduce loss maps in presence of collimators and other aperture restrictions.
The \st code \cite{st_man} was originally developed to study dynamic aperture of circular machines and non--linearities, and through the years it has been modified to include the possibility to track a large number of particles and interactions with collimators \cite{st_coll}.
The \st code provides a symplectic, fully chromatic and 6D tracking along the magnetic lattice of circular accelerators, taking into account interaction with obstacles (as collimators) and the machine aperture model.
The machine optic and aperture is produced with the \mad code~\cite{mad}. 
In \st, a crystal routine has been developed \cite{prev,cry_rout,dan_yellow,dan_th}.
This Monte--Carlo routine can describe the coherent interaction between charged particles and crystalline lattice, adapt the scattering routine introduced above for silicon crystals, and add a subroutine to evaluate ionisation energy loss.
The routine has been benchmarked with single and multi--pass experimental data. 

It has to be pointed out that BLM signals are produced by hadronic showers coming from lost particles interacting with the beam pipe, and all the other materials between the point where the proton is lost and the BLM position~\cite{Skordis2015Coupling}.
More detailed simulation that take as input to energy deposition tools the distribution of lost proton are also routinely performed to assess the LHC collimation performance, but this is beyond the scope of this work.
In \st is possible to produce the beam loss pattern, i.e. the expected number of proton lost per meter on the machine aperture.
This means that the loss pattern and the BLM signals can be deemed approximately comparable, if one consider the loss distribution patterns to be equal in both measurements and simulations.
This allows to perform relative estimations between the simulation and the loss maps measurements.

Loss maps of crystal collimation cleaning with proton beam at \SI{6.5}{\tera\electronvolt}, with the same settings presented in Tab. \ref{tab:coll_p}, were simulated using the crystal simulation routine.
The loss patterns can be compared to assess if the crystal collimation setup is well reproduced in simulations.
In Fig.~\ref{fig:b1_lmch_p} the measured Loss Maps are paired to the simulated loss patterns.
In Fig.~\ref{subfig:lm_std_h} and~\ref{subfig:lm_std_h_sim} the standard collimation LM and the simulated loss pattern are displayed for the full LHC ring.
The same loss patterns are presented with a magnification in the collimation insertion (IR7) in Fig.~\ref{subfig:lm_std_h_ir7} and~\ref{subfig:lm_std_h_ir7_sim}.
The same set of plot is shown for crystal collimation, for which the horizontal Cfg\#1 from Tab.~\ref{tab:coll_p} is chosen as an example.
In Fig.~\ref{subfig:lm_cry_h} and~\ref{subfig:lm_cry_h_sim} the full ring loss patterns are displayed, while Fig.~\ref{subfig:lm_cry_h_ir7} and~\ref{subfig:lm_cry_h_ir7_sim} shows the magnification in the IR7 region.

In the full ring it is possible to acknowledge how the simulations can reproduce, qualitatively very well, the main loss peaks observed during measurements.
In the IR7 magnifications (see Fig.~\ref{subfig:lm_cry_h_ir7} and~\ref{subfig:lm_cry_h_ir7_sim}), the loss pattern at the crystal location is as high as the one at the TCSG.
This is a feature of the \st code, which normalise proton losses to one meter, while the crystal is \SI{4}{\milli\meter} long.
This means that the loss value at the crystal location has to be reduced by a factor \num{250}.
By analysing the two loss patterns, it is evident that both the presence of collimators upstream and the use of the secondary collimators and absorbers are well reproduced in simulations.
Losses in Q7 cannot be directly compared; in simulations, no particle is lost in that region (for both standard and crystal system), because no off--momentum loss can be achieved (see Fig.~\ref{fig:disp_adv_ir7}) and hadronic showers are not accounted for in \st.

\begin{table*}[t!]
\caption{Collimation leakage ratio (standard vs. crystal) for proton beams. Measurements and simulations leakages are reported for each layout listed in Tab.~\ref{tab:coll_p}.}
\label{tab:lm_p}
\centering
\begin{tabular}{lcccccccccccc}
\hline\noalign{\smallskip}
    				&\multicolumn{10} {c} {Leakage Ratio} \\
    				&\multicolumn{6}{c}{IR7--DS}                                                                    &\multicolumn{2}{c}{IR3}  &\multicolumn{2}{c}{IR6}\\
Config.				&\multicolumn{2}{c}{Q7}			&\multicolumn{2}{c}{Q8--9}		&\multicolumn{2}{c}{Q10--11}    & \multicolumn{2}{c}{ }         & \multicolumn{2}{c}{}\\
                    & Meas. & Sim.                  & Meas. & Sim.                  & Meas. & Sim.                  & Meas. & Sim.                  & Meas. & Sim.\\
\hline\noalign{\smallskip}
H--1	 	& \num{ 0.34 \pm 0.06 } & --& \num{ 2.34 \pm 0.53 } &\num{9.99} & \num{ 1.54 \pm 0.40 } &\num{7.74} & \num{ 1.80 \pm 0.33 } &\num{89.76} & \num{ 0.30 \pm 0.06 }  &\num{1.39}\\
H--2		& \num{ 0.56 \pm 0.08 } & --& \num{ 3.67 \pm 0.59 } &\num{11.46}& \num{ 2.60 \pm 0.44 } &\num{9.15} & \num{ 2.65 \pm 0.42 } &\num{56.07} & \num{ 0.56 \pm 0.09 }  &\num{1.38}\\
H--3		& \num{ 0.63 \pm 0.11 } & --& \num{ 3.15 \pm 1.10 } &\num{11.36}& \num{ 2.07 \pm 1.11 } &\num{8.90} & \num{ 2.87 \pm 0.81 } &\num{49.70} & \num{ 0.49 \pm 0.12 }  &\num{1.19}\\
H--4		& \num{ 0.27 \pm 0.06 } & --& \num{ 2.99 \pm 0.53 } &\num{12.87}& \num{ 2.03 \pm 1.09 } &\num{11.04}& \num{ 2.07 \pm 0.44 } &\num{23.74} & \num{ 0.33 \pm 0.07 }  &\num{1.25}\\
H--5		& \num{ 0.07 \pm 0.01 } & --& \num{ 0.58 \pm 0.05 } &\num{12.65}& \num{ 1.73 \pm 0.24 } &\num{11.14}& \num{ 2.37 \pm 0.22 } &\num{29.44} & \num{ 0.02 \pm 0.01 }  &\num{0.22}\\
H--6		& \num{ 0.07 \pm 0.01 } & --& \num{ 0.64 \pm 0.07 } &\num{11.80}& \num{ 2.20 \pm 0.50 } &\num{10.44}& \num{ 2.37 \pm 0.27 } &\num{23.14} & \num{ 0.03 \pm  0.01 } &\num{0.22}\\
V--1		& \num{ 3.49 \pm 1.54 } & --& \num{ 16.43 \pm 9.60 }&\num{20.00}& \num{ 11.25 \pm 2.99 }&\num{22.45}& \num{31.05 \pm 21.53} &\num{34.76} & \num{ 1.20 \pm 0.65 }  &\num{11.61}\\
V--2		& \num{ 0.17 \pm 0.01 } & --& \num{ 3.34 \pm 0.27 } &\num{12.33}& \num{ 3.17 \pm 0.12 } &\num{14.47}& \num{ 4.08 \pm 0.29 } &\num{15.96} & \num{ 0.03 \pm 0.01 }  &\num{0.03}\\
V--3\text{*}& \num{ 0.42 \pm 0.06 } & --& \num{ 0.68 \pm 0.10 } &\num{0.40} & \num{ 0.73 \pm 0.07 } &\num{0.51} & \num{ 0.62 \pm 0.09 } &\num{0.30}	 & \num{ 0.08 \pm 0.01 }  &\num{0.45}\\
\hline\noalign{\smallskip}
\end{tabular}
\footnotesize{\text{*}Crystal oriented in amorphous.}
\end{table*}

\section{Cleaning Measurements and Simulations Discussion}\label{comparison}

To asses performance of crystal collimation, the ratio of the leakage factors measured with the standard collimation to the one measured with crystal system is considered.
In any given region, a ratio larger than \num{1} indicates an improved cleaning performance of the crystal collimation system with respect to the standard one.
The focus is on the \emph{leakage ratios} measured in IR7-DS (in the three different regions defined in Section~\ref{sec:lossmaps}), on the momentum cleaning primary collimator (TCP IR3), and on collimators in IR6 region; detail are given in Table~\ref{tab:lm_p}.

In simulations, the cleaning efficiency is evaluated by averaging the cold losses in the same regions of interest defined in measurements. 
The simulated cleaning ratios, between standard and crystal collimation, are reported in Tab.~\ref{tab:coll_p}.

\begin{SCfigure}[][tp]
\centering
\includegraphics[width=0.5\linewidth]{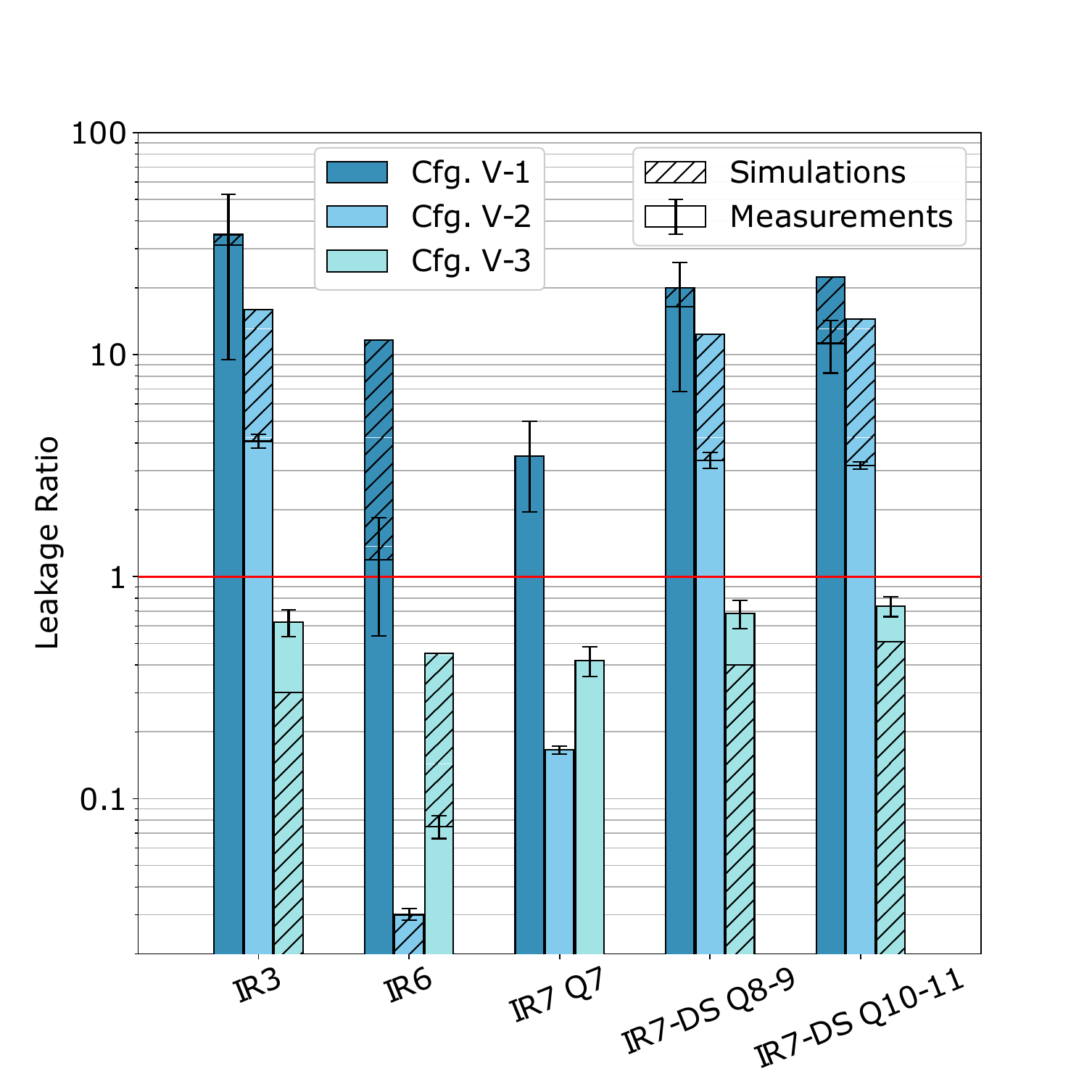}
\caption{\protect\rule{0ex}{9ex}Leakage ratio with respect to standard collimation in several LHC location, for proton beams at top energy. Measurements and simulations are shown by solid and dashed bars, respectively, for the vertical configurations in Tab.~\ref{tab:coll_p}.
The red line show the standard collimation reference: above the line the crystal collimation is improving the cleaning performance of the present system.}
\label{fig:lmv_p_sim} 
\end{SCfigure}



\subsection*{Vertical Plane}

For the vertical crystal Cfg\#1 an improvement by a factor \num{3} is observed in all the regions defined in Section~\ref{sec:lossmaps}. 
Moreover a loss reduction of a factor higher than \num{10} is observed in both Q8--9 and Q10-11 regions (IR7--DS).
Also, on IR3 TCP losses are lower by a factor \num{\sim 30}.
When a reduced set of TCSGs is used in Cfg\#2, 
reduced performance are registered, as shown in Fig.~\ref{fig:lmv_p_sim}.
Finally, when orienting the crystal as an amorphous material, one observes a general worsening in the system performance. 
Furthermore, the crystal collimation becomes between \SIrange{30}{50}{\percent} less performing than the standard collimation in every region under investigation.

In simulations, the relative performance with respect to standard collimation is reasonably well reproduced, as shown in Fig.~\ref{fig:lmv_p_sim}, apart for the observation of losses in Q7 region.
In particular, for Cfg\#1, simulation expectations are within the measurement error bars, except for the IR6 cluster.
In the simulation is expected to be \num{10} times better than standard collimation, while it is observed to have the same performance.
In Cfg\#2 where only TCLAs are used\footnote{at an aperture of \SI{11}{~\sig}} to absorb the channeled halo, crystal collimation performance is \num{2} times more efficient in simulation than in measurements, in IR3 and the IR7--DS.
Finally, the amorphous orientation Cfg\#3 presents leakage ratios in agreement between simulation and measurements.

\begin{figure}[tp!]
\centering
\begin{subfigure}[t]{.47\textwidth}
\centering
\includegraphics[width=\linewidth]{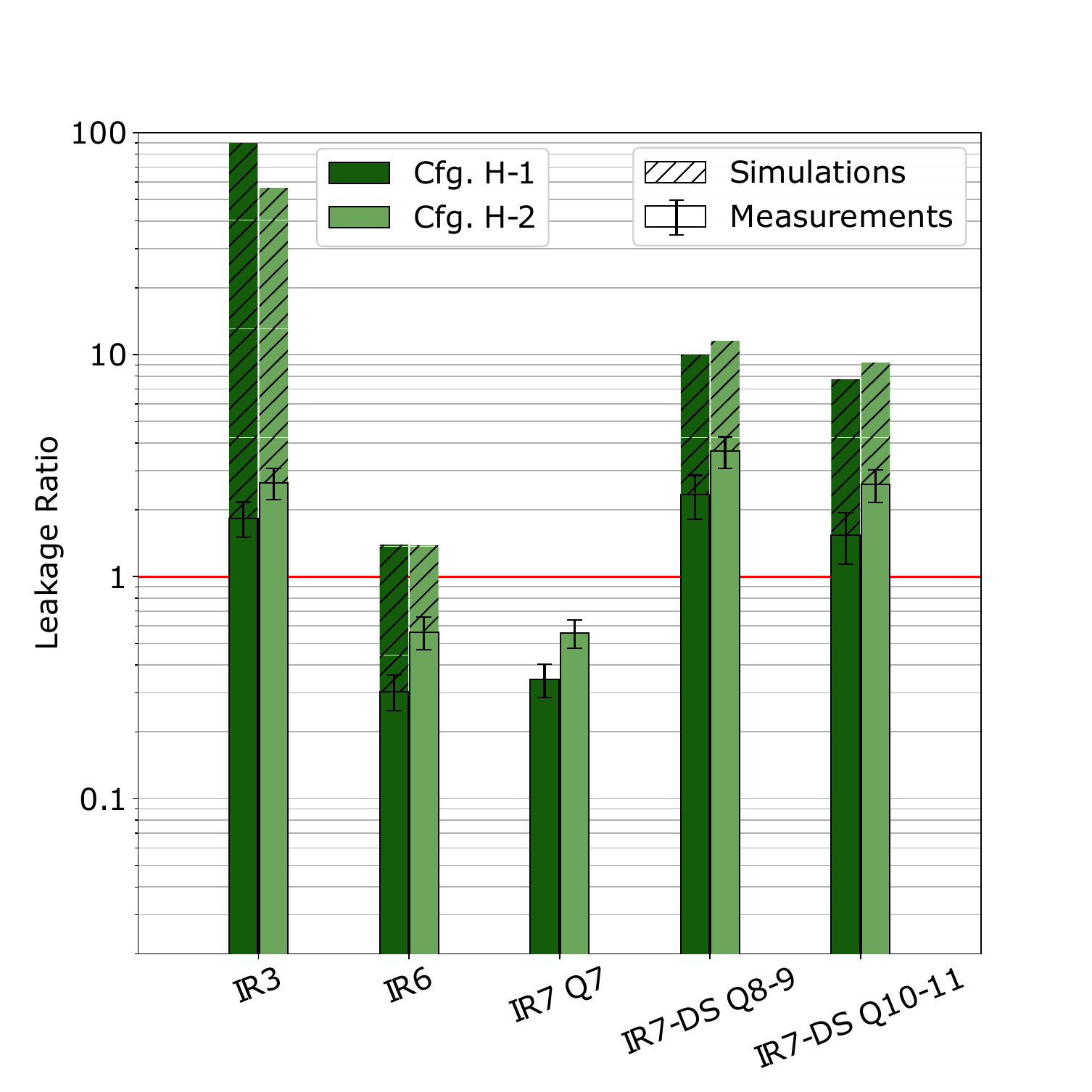}
\caption{Configuration \#1, \#2 . Comparison for upstream collimators effect.}\label{subfig:lm_h_comp_up}
\end{subfigure}
\begin{subfigure}[t]{.47\textwidth}
\centering
\includegraphics[width=\linewidth]{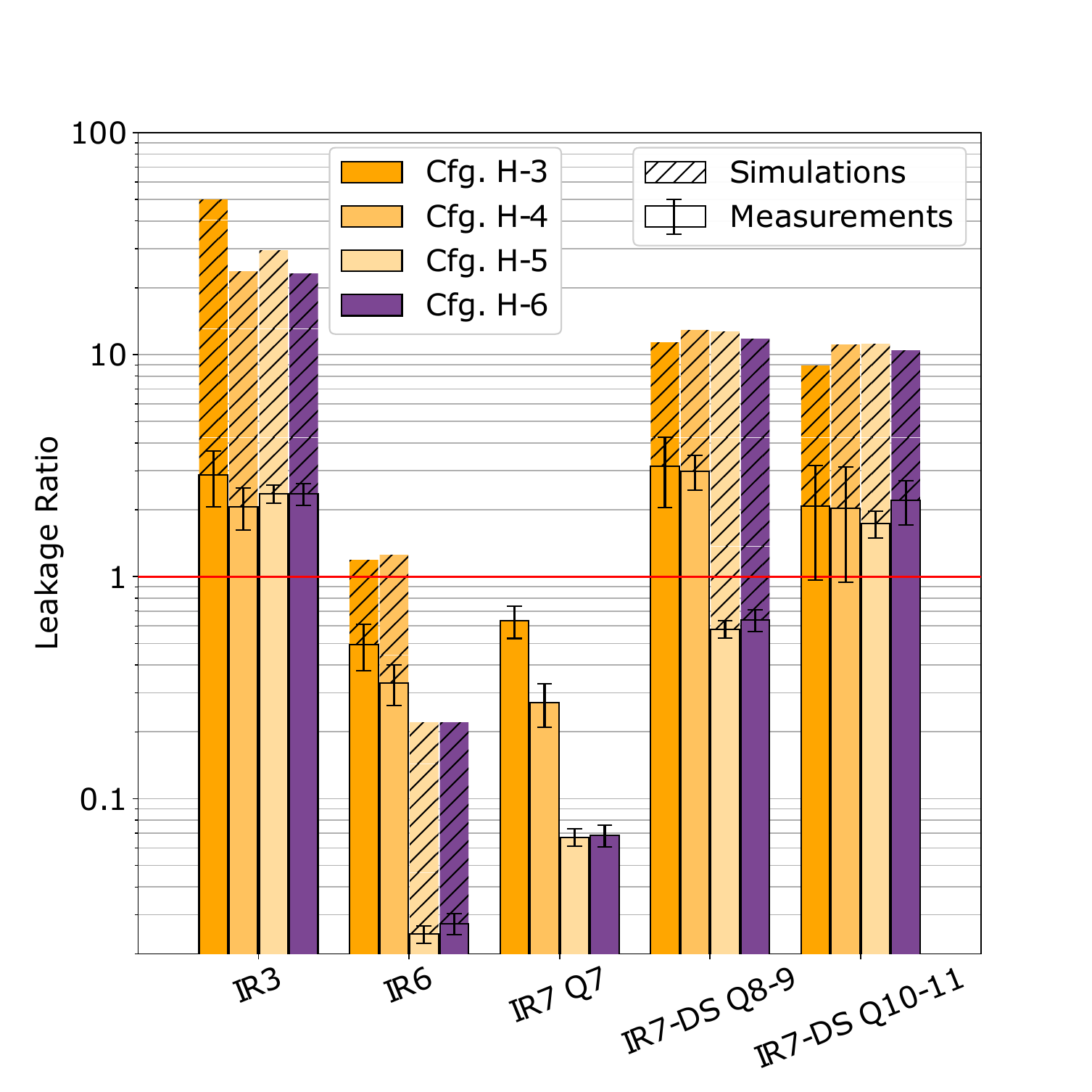}
\caption{Configuration \#3, \#4 and \#5, \#6. Comparison for retraction of collimators downstream the first TCSG used as absorber and the minimal reduced settings for horizontal layout.}\label{subfig:lm_h_comp_dw}
\end{subfigure}
\caption{Leakage ratio with respect to standard collimation in several LHC location, for proton beam at top energy. 
Measurements and simulations are shown by solid and dashed bars, respectively, for the horizontal configurations in Tab.~\ref{tab:coll_p}.
The red line show the standard collimation reference: above the line the crystal collimation is improving the cleaning performance of the present system.}
\label{fig:lmh_p_sim} 
\end{figure}

\subsection*{Horizontal Plane}
In the horizontal plane, a worsening or a matching of cleaning performance is observed with any configuration of crystal collimation setup.
In particular, what appears to consistently limit the performance is the leakage in Q7 (see Figs.~\ref{fig:lmh_p_sim}).
Comparing \#1 (Cfg\#1) and \#2 in the horizontal plane, demonstrates that the presence of upstream collimators reduces the performance of the crystal system of about \SI{\sim40}{\percent} in any region in consideration, as shown in Fig.~\ref{subfig:lm_h_comp_up}.
Dedicated studies were performed in SPS with the UA9 crystal collimation experimental setup, quantitatively observing a worsening of the crystal channeling performance in presence of upstream collimators~\cite{rossi_phd}. 
In Fig.~\ref{subfig:lm_h_comp_dw} the leakage ratios measured in Cfg\#3, \#4 and \#5 are presented and clarify that removing secondary collimators downstream worsen the system performance in the dispersion suppressor.
this can be seen as a demonstration of the not sufficient stopping power of the TCSG to absorb the channeled halo, therefore several of them are needed to obtain a reduction of losses in the DS.
Additionally, the absence of the first horizontal TCSG in Cfg\#5 and \#6, reduces drastically the absorbing capacity of the system. 
In fact the worse performance are observed in both Q7 and IR6\footnote{This location is, in fact, on the proper phase advance with respect to the first TCSG, so that all particles escapes IR7 to be lost in IR6, after almost a full revolution of the machine.} and on the first dispersion peak.
Conversely, due to the reduced amount of material, the production of off-momentum particles is decreased and good leakage ratios are observed in IR3.


In~\cite{dech_lhc} it is shown how the horizontal crystal has a curvature radius close to the critical value for which channeling does not appear, and how this feature enhances the dechanneled particles at low deflection angle.
Thus, with respect to the vertical crystal, more particles leave the crystal with trajectories closer to the beam envelope, and not properly deflected.
Secondary collimators placed at transversal aperture larger than the crystal are only able to intercept the halo deflected of a certain angle~\cite{dan_layout}.
Given a fixed aperture setup for crystals and secondary collimators, the range of the angular cut vary with the obstacle location and its optic functions.
This might explain how the absence of the first of the two secondary horizontal collimators (that with the same transversal aperture can intercept halo particles at lower deflection), or the absence of a proper screen by the downstream secondary skews, enhances losses on Q7.   
Additionally, the trend observed in both horizontal and vertical measurement, leads to the hypothesis that losses in the Q7 region are produced by hadronic showers produced in the very upstream TCLAs.


An overall factor \num{\sim 10} of improvement with respect to standard collimation is expected in simulations in almost every configuration for DS cleaning ratio, as shown in Fig.~\ref{fig:lmh_p_sim}.
While other configuration seems to have the same relative difference with respect to the theoretical expectations, this is not observed for the leakage in Cfg\#5 and Cfg\#6.
In particular, the cluster Q8--9, that in simulations is expected to be at the same level of the other configurations, while in measurements it is observed almost one order of magnitude lower.
This feature might be explained by the same hypothesis made for Q7 cluster, i.e. the absence of proper absorption to stop the showers coming from both the TCSGs and the TCLAs.
Due to the absence of most of TCSGs in these configurations, single diffractive protons coming from the collimators might reach the cluster Q8--9 and dope the measured leakage.
This would be a confirmation of an observation already done in the Super Proton Synchrotron~\cite{sps_loss}, where losses on the first dispersive peak comes from off-momentum particles emerging from collimators, while losses on the second dispersive peak comes from off-momentum particles that have only interacted with the crystal.

\subsection*{Simulation Comparison Remarks}

In simulations, the expected collimation performance of horizontal and vertical crystals are closer than in measurements. 
To explain the difference observed between the horizontal and vertical plane, one should recall again that the horizontal crystal is significantly closer to the critical radius (at \SI{6.5}{\tera\electronvolt}) with respect to the vertical one~\cite{dech_lhc}.
It is known and described in \cite{dan_th}, that the absence of an analytic model that describes dechanneling makes still not possible to simulate properly crystals with curvature few times larger than the critical radius. 

To reproduce the real cleaning performance of crystal collimation, and the full BLM loss pattern measured in the LMs, a \fl simulation with energy deposition and showers tracking is needed and can be performed using the presented loss pattern produced by \st as an input.
This study will be fundamental to verify the hypothesis made in this work.

\section{Conclusions}\label{sec:concl}
A systematic study, to compare the cleaning performance of crystal collimation has been measured and compared to that of standard collimation system, was carried out. 
Both system were probed starting from the same nominal configurations~\cite{martinez2019run}.
In the horizontal plane, a slight improvement of cleaning is observed in the dispersion suppressor clusters at the critical locations with highest losses for the standard system. 
With the vertical plane a good improvement is observed in all DS regions by a factor \SIrange[]{\sim3}{\sim10}{}, and in general along the whole ring.
The different behaviour of horizontal and vertical crystals might be explained by the fact that the radius of the horizontal crystal is \SI{20}{\percent} smaller than that of the vertical one, hence the deflection efficiency of the \SI{6.5}{\tera\electronvolt} halo particles is considerably larger for the vertical than for the horizontal crystal collimation system.

With the crystal collimation system a different source of inefficiency is located at the very first cold quadrupole downstream the collimation insertion.
In this region, the dispersion is very small, hence off-momentum losses should be negligible.
The hypothesis made upon these observations is that losses are generated by direct hadronic showers coming from the close upstream TCLAs.

Measurements have shown that collimators upstream the crystal aperture reduce the performance of the system.
This was later confirmed in dedicated studies in the CERN SPS with the UA9 scaled crystal collimation experimental setup.
It has also been observed that the deflected halo is not completely absorbed by secondary collimators. 
When inserting the full set of collimators downstream the crystal, the deflected particles are better absorbed and the crystal collimation performance is improved.

Particle tracking simulations were used to compute the expected beam loss pattern and to compare it to the measured ones.
While the computed loss pattern is in a fair agreement with data, the measured leakages are still not directly comparable with simulations.

A comprehensive simulation with the geometry of the whole collimation insertion (with all devices and monitors), accounting for the energy deposition and hadronic showers evolution, should be used for a quantitative benchmark of the measured leakages and in particular to test the hypothesis made on the losses observed in the Q7 region.

\section*{Acknowledgement}
The authors would like to acknowledge all the CERN teams and groups who contributed to the experimental measurements, in particular the ABP group and OP group in the Beams department and the STI and SMM groups in the Engineering department. A special acknowledgement goes to the members of the UA9 collaboration, in particular to the INFN team and the PNPI team who built the crystals installed in the LHC. This work was supported by the HL-LHC project and by the Collimation project at CERN.
%
%

\bibliography{biblio.bib}

\end{document}